\documentclass[%
	reprint,
	superscriptaddress,
	amsmath,amssymb,amsfonts,
	aps,
	pra,
]{revtex4-2}

\usepackage[utf8]{inputenc}
\usepackage[T1]{fontenc}
\usepackage{lmodern}
\usepackage[english]{babel}
\usepackage[autostyle=true]{csquotes}

\usepackage[dvipsnames]{xcolor}
\definecolor{BLUE}{named}{blue}
\usepackage{graphicx}
\graphicspath{{figures/}}

\usepackage{tikz}
\usepackage{tikz-cd}
\usepackage[normalem]{ulem}
\usepackage{mhchem}
\usetikzlibrary{%
  matrix,%
  calc,%
  arrows%
}

\usepackage{amssymb,amsmath}
\usepackage{bm}
\usepackage{mathtools}
\usepackage{braket}

\usepackage{xspace}
\usepackage{mleftright}
\mleftright

\usepackage{booktabs}
\usepackage{tabularx}
\newcolumntype{L}{>{\raggedright\arraybackslash}X}
\newcolumntype{Y}{>{\centering\arraybackslash}X}
\newcolumntype{R}{>{\raggedleft\arraybackslash}X}

\usepackage{hyperref}
\hypersetup{
}

\usepackage[%
    separate-uncertainty=false,
    range-phrase=-,
]{siunitx}

\DeclareSIUnit\bohr{\text {\ensuremath {a}}_{0}}

\usepackage[
	capitalise,
]{cleveref}

\definecolor{mypurple}{rgb}{0.49,0.18,0.56}
\definecolor{mygold}{rgb}{0.93,0.49,0.13}
\definecolor{mygreen}{rgb}{0,0.5,0}
\definecolor{myblue}{rgb}{0,0,0.75}
\definecolor{mymagenta}{cmyk}{0,1,0,0.12}
\definecolor{mygray}{rgb}{0.5,0.5,0.5}

\newif\ifcomments
\commentstrue 

\newcommand*{\etothepowerof}[1]{\mathrm{e}^{#1}}
\newcommand*{\etothe}[1]{\etothepowerof{#1}}
\newcommand*{\bigO}{\mathcal{O}}

\renewcommand{\vec}{\bm}

\begin{document}
\title{The role of higher-order terms in trapped-ion quantum computing with magnetic gradient induced coupling}

\author{Sebastian Nagies}
\email{sebastian.nagies@unitn.it}
\thanks{Corresponding author}
\affiliation{Pitaevskii BEC Center and Department of Physics, University of Trento, Via Sommarive 14,
38123 Trento, Italy}
\affiliation{INFN-TIFPA, Trento Institute for Fundamental Physics and Applications, Trento, Italy}

\author{Kevin T. Geier}
\affiliation{Pitaevskii BEC Center and Department of Physics, University of Trento, Via Sommarive 14,
38123 Trento, Italy}
\affiliation{INFN-TIFPA, Trento Institute for Fundamental Physics and Applications, Trento, Italy}
\affiliation{Quantum Research Center, Technology Innovation Institute, P.O. Box 9639, Abu Dhabi, United Arab Emirates}

\author{Javed Akram}
\affiliation{eleQtron GmbH, Heeserstr. 5, 57072 Siegen, Germany}

\author{Junichi Okamoto}
\affiliation{eleQtron GmbH, Heeserstr. 5, 57072 Siegen, Germany}

\author{Dimitrios Bantounas}
\affiliation{eleQtron GmbH, Heeserstr. 5, 57072 Siegen, Germany}

\author{Christof Wunderlich}
\affiliation{eleQtron GmbH, Heeserstr. 5, 57072 Siegen, Germany}
\affiliation{Department of Physics. School of Science and Technology, University of Siegen, 57068 Siegen, Germany}

\author{Michael Johanning}
\affiliation{eleQtron GmbH, Heeserstr. 5, 57072 Siegen, Germany}

\author{Philipp Hauke}
\email{philipp.hauke@unitn.it}
\thanks{Corresponding author}
\affiliation{Pitaevskii BEC Center and Department of Physics, University of Trento, Via Sommarive 14,
38123 Trento, Italy}
\affiliation{INFN-TIFPA, Trento Institute for Fundamental Physics and Applications, Trento, Italy}

\begin{abstract}
Trapped-ion hardware based on the Magnetic Gradient Induced Coupling (MAGIC) scheme is emerging as a promising platform for quantum computing. 
Nevertheless, in this---as in any other---quantum-computing platform, many technical questions still have to be resolved before large-scale and error-tolerant applications are possible.  
In this work, we present a thorough discussion of the structure and effects of higher-order terms in the MAGIC setup, which can occur due to anharmonicities in the external potential of the ion crystal (e.g., through Coulomb repulsion) or through curvature of the applied magnetic field. 
These terms generate systematic shifts in the leading-order interactions and take the form of three-spin couplings, two-spin couplings, local fields, as well as diverse phonon-phonon conversion mechanisms.  
We find that most of these are negligible in realistic situations, with only two contributions that need careful attention. First, there are undesired longitudinal fields contributing shifts to the resonance frequency, whose strength increases with chain length and phonon occupation numbers; while their mean effect can easily be compensated by additional $Z$ rotations, phonon number fluctuations need to be avoided for precise gate operations.
Second, anharmonicities of the Coulomb interaction can lead to well-known two-to-one conversions of phonon excitations. Both of these error terms can be mitigated by sufficiently cooling the phonons to the ground-state. 
Our detailed analysis constitutes an important contribution on the way of making magnetic-gradient trapped-ion quantum technology fit for large-scale applications, and it may inspire new ways to purposefully design interaction terms.
\end{abstract}

\maketitle

\section{Introduction}\label{sec:intro}

The rapid development of quantum-computing devices promises a significant technological impact, with application fields ranging from pharmacology over material design to fundamental research into the building blocks of nature \cite{Ladd2010, Fedorov2022, Yarkoni2022, Beck2023, Hoefler2023, Scholten2024, Troyer2024}. A highly promising approach for trapped-ion based quantum computing is the Magnetic Gradient Induced Coupling (MAGIC) scheme, which utilizes a magnetic field gradient to couple ions and make them individually addressable with microwaves  
\cite{Mintert2001, Wunderlich2002, Johanning2009, Zippilli2014, Piltz2016, Bassler2023, Weidt2016, Arrazola2018, Leu2023}. 
It integrates some of the major strengths of traditional trapped-ion
platforms \cite{Wineland2003,  Leibfried2003a,HAFFNER2008,Bruzewicz2019,Srinivas2021}, including extended qubit lifetimes, minimal interference, and all-to-all connectivity.
An advantage of the magnetic-gradient induced coupling scheme is its negligible crosstalk \cite{Piltz2014} and the high stability of microwave fields, in addition to the possibility to integrate microwave technology into trap chips. 
However, as for any quantum computing platform on the way to becoming a mature technology \cite{Preskill2018,Cai2023}, various engineering and fundamental problems still need to be addressed, in particular what concerns the role of imperfections and errors. 

\begin{figure}
    \centering
    \includegraphics[width=\columnwidth]{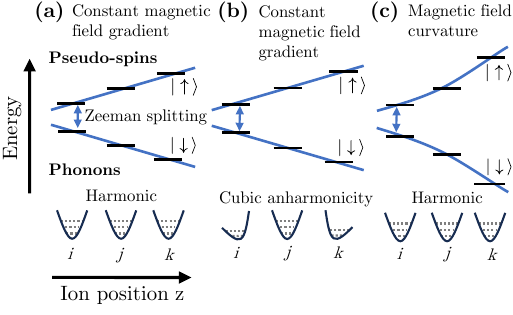}
    \caption{\label{fig:higherorder_sources}%
    $\mathbf{(a)}$ The Magnetic Gradient
Induced Coupling (MAGIC) scheme for trapped-ion quantum information processing employs a magnetic field gradient, which renders the resonance frequency of the internal pseudo-spin position-dependent. This couples the spin to the spatial phonon degree of freedom, which effectively generates a spin--spin coupling that can be used for entangling gates. 
    In this work, we analyse higher-order contributions in such a setup, in particular  
    $\mathbf{(b)}$ anharmonicities in the vibrational degrees of freedom due to Coulomb repulsion and 
    $\mathbf{(c)}$ curvature of the magnetic field, which can lead to additional three-body and two-body interactions, local field corrections, as well as spin--phonon couplings.}
\end{figure}

In this work, we investigate the effects of higher-order contributions in the MAGIC setup (illustrated in \cref{fig:higherorder_sources}). 
These include in particular residual terms originating from a third-order expansion of the Coulomb interactions (see also~\cite{Loewen2003}), spatial dependence of the magnetic field gradient, and anharmonicities of the individual trapping potentials. 
While our discussion focuses mainly on systematic shifts of spin interaction strengths and local fields, we also indicate how they apply to discrete quantum gates. 
As we show, applying the canonical unitary transformation on the axial phonon modes that is used to derive the MAGIC scheme for trapped-ion logic \cite{Mintert2001, Wunderlich2002, Johanning2009}, such higher-order terms induce three-qubit interactions as well as various types of terms that couple the pseudo-spins and the phonon vibrations. We show that in realistic situations the strength of most of these terms is negligible. 
Two contributions, however, can naturally achieve considerable strengths, especially when scaling to large ion chains. The first one takes the form of a phonon-occupation-dependent longitudinal field, whose strength increases with ion chain length and which, for typical parameters, can become comparable to the native two-qubit interaction strength. While the mean effect of such terms can easily be compensated by detuning the microwave frequency from the qubit resonance, virtual $Z$~rotations~\cite{McKay2017}, or through suitable dynamical decoupling schemes \cite{Cai2023}, phonon number fluctuations lead to residual field fluctuations that require careful cooling techniques to avoid moving off resonance or---in the case of discrete gates---to maintain gate fidelity.
The second type of contribution that needs care is a (well-known) two-to-one conversion between phonon modes \cite{Marquet2003}. These vanish if all phonon modes are ground-state cooled. 
As three-qubit interactions and spin--phonon couplings may be desired terms for various applications, ranging from quantum simulation \cite{Yang2016,GonzalezCuadra2018, GonzalezCuadra2019,Andrade2022,Mildenberger2022} to quantum optimization \cite{Lechner2015,Cepaite2023,Nagies2025a}, we also discuss possible strategies to engineer such terms.  
This work constitutes a systematic analysis of higher-order terms in the MAGIC setup, an important step for transitioning this type of quantum hardware from academic prototypes to mature technology. 
We also hope that it may inspire similar studies in the context of other platforms, such as trapped ions with laser-free entangling gates generated by oscillating magnetic fields \cite{Ospelkaus2008, Ospelkaus2011, Yu2022,Loeschnauer2024}, or superconducting qubits \cite{Blais2007}.

This paper is structured as follows. 
In Sec.~\ref{sec:magic}, we review the MAGIC scheme for trapped-ion quantum information processing. 
In Sec.~\ref{sec:3body}, we compute the influence of terms due to various third-order expansions, including numerical estimates of their resulting strengths. These are a third-order expansion of the Coulomb repulsion between ions (Sec.~\ref{subsec:coulomb}) and an external anharmonicity acting locally at each ion (Sec.~\ref{subsec:cubicexternal}). Sec.~\ref{subsec:higherorderB} is dedicated to the effects of an applied magnetic field with non-vanishing curvature. We compare the resulting three-body interaction strengths to different schemes known in the literature to generate such terms in Sec.~\ref{subsec:3body}.
Finally, Sec.~\ref{sec:conclusion} contains our conclusions.

\section{Quantum information processing with trapped ions subject to a magnetic field gradient}\label{sec:magic}
The MAGIC  scheme for quantum computation \cite{Mintert2001,Wunderlich2002,Johanning2009} uses microwaves instead of optical lasers to address ions and implement quantum gates. Since microwaves cannot be focused on individual ions, a magnetic field gradient is used to Zeeman-shift the resonance frequency of each ion and make them individually addressable in frequency space. Furthermore, the magnetic field gradient induces an Ising-type always-on all-to-all coupling between the ions, even in the absence of any driving fields, which can be used to synthesize quantum gates \cite{Khromova2012,Bassler2023} or perform adiabatic quantum sweeps \cite{Huber2021}. Here, we give a concise review of the derivation of the spin--spin coupling. We refer to the literature for further details \cite{Mintert2001,Wunderlich2002,Johanning2009}.

\subsection{Effective spin--spin interaction}

We consider a one-dimensional (1D) ion chain, elongated in the $z$ direction.
The ions are subject to a static magnetic field with a constant gradient~$\partial_z B$ in the axial direction $z$, $\vec{B}(z) = \vec{B}_0 + \mathinner{\partial_z B} z\vec{\hat{e}}_z + \bigO(z^2)$, where $\vec{B}_0$ is a constant offset field.
In Appendix~\ref{app:transversal}, we generalize this setting to gradients in the transversal directions. Moreover, we will discuss the effect of a field with non-vanishing curvature in Sec.~\ref{subsec:higherorderB}.
We further assume that we use magnetically sensitive states as our qubit states (typically states from a hyperfine ground state manifold with a resonance frequency in the microwave regime \cite{Balzer2006,Piltz2016, Weidt2016}) to make them individually addressable.
Although such states are typically prone to dephasing caused by magnetic field fluctuations, long coherence times can be achieved via suitable protection schemes such as dynamical decoupling~\cite{Piltz2013, Cai2023} or microwave dressing~\cite{Timoney2011}. In Ref.~\cite{Ruster2016}, coherence times of $2.1(1)$ seconds are demonstrated for electron spin states in a single trapped \ce{^{40}_{}Ca+} 
ion using effective magnetic shielding techniques.

The spatially varying magnetic field gives rise to a position-dependent resonance frequency~$\omega(z)$, which in general has to be calculated numerically or can be estimated using the Breit--Rabi formula~\cite{Breit1931}.
Under weak Zeeman splitting, which holds for the examples considered in this work, provided that the offset field~$\vec{B}_0$ is sufficiently small, the resonance frequency can be obtained as $\omega(z) = \omega_0 +  (g_F^{(e)}m_F^{(e)} -  g_F^{(g)}m_F^{(g)}) \mu_{\mathrm{B}} B(z) /\hbar$.
Here, $g_F^{(g)}$ and $g_F^{(e)}$ are, respectively, the Landé $g$-factors of the qubit ground and excited state with total atomic angular momentum~$F$, $m_F^{(g)}$ and $m_F^{(e)}$ are the corresponding magnetic quantum numbers, and $\mu_{\mathrm{B}}$ is the Bohr magneton.
The bare resonance frequency in the absence of a magnetic field is denoted by $\omega_0$.
The internal energy of the ions is then given by
\begin{align}
H_{\mathrm{int}} = -\frac{\hbar}{2} \sum_{n=1}^N \omega(z_n) \sigma_Z^{(n)} \,,
\end{align}
where $\sigma_Z^{(n)}$ denotes the Pauli-$Z$~operator acting on qubit~$n$ at position~$z_n$ and $N$ is the number of ions in the chain.

The resonance frequency can be expanded around the equilibrium position $z_{n,0}$ of ion $n$ as
\begin{align}
    \label{eq:resonancefrequency_gradient}
    \omega(z_n) = \omega_n + \partial_z \omega_n (z_n - z_{n,0}) + \bigO(z_n^2) \,,
\end{align}
where $\omega_n=\omega(z_{n,0})$ is the resonance frequency and $\partial_z \omega_n = \partial \omega(z_{n,0}) / \partial z$ the first derivative at the position of the ion.
Higher-order terms in this expansion will be discussed in Sec.~\ref{subsec:higherorderB}.
Denoting the excursion of ion $n$ from its equilibrium position as $q_n = z_n - z_{n,0}$, we can then write the internal energy as
\begin{align}
    \label{eq:Hint}
    H_{\mathrm{int}} = - \frac{\hbar}{2} \sum_{n=1}^N \omega_n \sigma_Z^{(n)}- \frac{\hbar}{2} \sum_{n=1}^N \partial_z \omega_n q_n\sigma_Z^{(n)}.
\end{align}

Additionally, the ions have an energy associated with their external degrees of freedom, composed of their kinetic energy, the potential energy due to the trap, as well as their mutual Coulomb repulsion:
\begin{align}
    \label{eq:Hext}
    H_{\mathrm{ext}} &= H_{\mathrm{kin}} + V(\vec{z})\nonumber\\
    &= H_{\mathrm{kin}} + V_\mathrm{trap}(\vec{z}) + V_\mathrm{Coul}(\vec{z})\\
     &= \frac{1}{2m} \sum_{n=1}^N p_{z,n}^2 + \frac{m}{2} \omega_z^2 \sum_{n=1}^N z_n^2 + \frac{e^2}{8\pi\epsilon_0} \sum_{n\neq l}^N \frac{1}{|z_n - z_l|} \,.\nonumber
\end{align}
Here, the vector $\vec{z}$ contains the axial positions of all ions, $e$ is the charge of the ions, $m$ their mass, $\epsilon_0$ the vacuum permittivity, and $p_{z,n}$ is the momentum of ion $n$ in the axial direction.
For now, we will consider a harmonic trap with angular frequency~$\omega_z$ and expand the potential energies around the ions' equilibrium positions up to second order, yielding (after dropping a constant term)
\begin{align}
H_{\mathrm{ext}} = \frac{1}{2m} \sum_{n=1}^N p_{z,n}^2 + \frac{m}{2} \sum_{n=1}^N \sum_{l=1}^N A_{nl} q_n q_l + \bigO(q^3) \,,
\end{align}
with $A_{nl} = \partial^2 V(\vec{z}_0) / \partial z_n \partial z_l$. 
In \cref{subsec:coulomb,subsec:cubicexternal}, we will discuss the effects of higher-order terms in the expansion of the potential energy.

The matrix $A$ is symmetric and real and can thus be diagonalized as $D = S^{-1} A S$ with an orthogonal matrix~$S$. Since $A$ is a positive-definite Hessian (it results from an expansion around the minimum of the potential energy), its eigenvalues are positive and can be written as $\nu_l^2$. We can now express the local coordinates $q_n$ in terms of collective normal modes $Q_l$, 
\begin{equation}
    \label{eq:normal modes}
    q_n = \sum_{l=1}^N S_{nl} Q_l \,.
\end{equation}
These diagonalize the external Hamiltonian to 
\begin{align}
\label{eq:uncoupled_harm}
    H_{\mathrm{ext}} = \frac{1}{2m} \sum_{n=1}^N P_{n}^2 + \frac{m}{2} \sum_{n=1}^N \nu_n^2 Q_n^2 \,,
\end{align}
where we have defined $\vec{P} = S^{-1} \vec{p}_z$ and only written down the second-order terms.

We can quantize the external Hamiltonian by introducing phonon creation and annihilation operators for the normal modes: 
\begin{equation}
    \label{eq:quantized modes}
    Q_l = \Delta z_l (a^\dagger_l + a_l) \,,
\end{equation}
with $\Delta z_l = \sqrt{\hbar / 2m\nu_l}$ being the width of the ground-state wave packet of the quantum harmonic oscillator. In terms of these operators, the external Hamiltonian becomes $H_{\mathrm{ext}} = \hbar \sum_{n=1}^N \nu_n a^\dagger_n a_n$ and
the overall Hamiltonian $H = H_{\mathrm{int}} + H_{\mathrm{ext}}$ can now be expressed as
\begin{align}
\label{eq:HMagic}
 H = &-\frac{\hbar}{2} \sum_{n=1}^N \omega_n \sigma_Z^{(n)} -\frac{\hbar}{2}\sum_{n=1}^N \sum_{l=1}^N \partial_z \omega_n  S_{nl} \Delta z_l (a^\dagger_l + a_l) \sigma_Z^{(n)} \nonumber\\
 &+ \hbar \sum_{n=1}^N \nu_n a^\dagger_n a_n \,.
\end{align}
 
We can decouple the internal and external degrees of freedom (spins and phonons) via a suitable unitary transformation~\cite{Mintert2001, Johanning2009} (in some contexts known as polaron transformation \cite{Xu2016}), $\Tilde{H} = U^\dagger H U$, with
\begin{subequations}
\label{eq:polarontrafo}
\begin{align}
    U &= \exp \left(-\frac{i}{\hbar}\sum_{n=1}^N\sum_{l=1}^N \Delta z_l \epsilon_{nl} \sigma_Z^{(n)} P_l\right) \,, \\
    \epsilon_{nl} &= \frac{\Delta z_l \partial_z \omega_n}{\nu_l} S_{nl} \,. \label{eq:lambdicke}
\end{align}
\end{subequations}
Here, we have defined the effective Lamb--Dicke parameter~$\epsilon_{nl}$, describing how strongly ion~$n$ (with internal transition frequency~$\omega_n$) couples to phonon mode $l$ (with frequency~$\nu_l$).

With the transformed phonon operators
\begin{align}
 U^\dagger a_l U =  a_l + \frac{1}{2}\sum_n \epsilon_{nl} \sigma_Z^{(n)}
\end{align}
(and unchanged spin operators $\Tilde{\sigma}_Z^{(n)} = \sigma_Z^{(n)}$), the total Hamiltonian yields the desired two-body spin--spin interaction,  
\begin{align}
\label{eq:magic_coupling}
    \Tilde{H}^{(2)} = &-\frac{\hbar}{2} \sum_{n=1}^N \omega_n \sigma_Z^{(n)} + \hbar \sum_{n=1}^N \nu_n a^\dagger_n a_n \nonumber\\
    &- \frac{\hbar}{2}\sum_{i<j}^N J_{ij}^{(2)} \sigma_Z^{(i)}\sigma_Z^{(j)} \,, 
\end{align}
with the spin--spin coupling strength
\begin{align} \label{eq:J2}
    J_{ij}^{(2)} = \sum_{n=1}^N \nu_n \epsilon_{in} \epsilon_{jn} \,.
\end{align}

This effective Hamiltonian is valid up to second order in the potential energy and linear order in the spatial dependence of magnetic field, owing to the Taylor expansions leading to the quadratic Hamiltonian in \cref{eq:HMagic}.
Note that the polaron transformation from Eq.~\eqref{eq:HMagic} to \cref{eq:magic_coupling} is exact in the sense that it does not involve any further approximations. 

The spin--spin coupling strength $J_{ij}^{(2)}$ can be tuned via the magnetic field gradient and the strength of the external harmonic trap, since $J_{ij}^{(2)} \sim (\partial_zB/\nu)^2$. By employing different trapping potentials along the chain (not necessarily harmonic), coupling strengths between different pairs of ions can be enhanced or reduced \cite{Zippilli2014}.

\subsection{\label{sec:discretemagicgates}Discrete quantum gates within the MAGIC scheme}

The always-on two-body interactions in the MAGIC setup, Eq.~\eqref{eq:magic_coupling}, can be used to implement entangling two-body quantum gates. To implement, for example, a unitary $ZZ$~rotation gate of the form $\exp(-i\theta \sigma_Z^{(i)} \sigma_Z^{(j)})$, one first transfers the state populations of all ions except $i$ and $j$ into states that are (to first order) magnetically insensitive. Consequentially, \cref{eq:magic_coupling} reduces in a rotating frame to
\begin{align}
\label{eq:magic_gate}
    \Tilde{H}^{(2)} = - \frac{\hbar}{2} J_{ij}^{(2)} \sigma_Z^{(i)}\sigma_Z^{(j)} \,. 
\end{align}
The system's free evolution for a time $T$ then implements the desired discrete unitary gate with a rotation angle of $\theta = - \frac{1}{2} J_{ij}^{(2)} T$. 
More complicated global entangling gates can also be synthesized by selectively excluding sets of ions from the two-body interactions and letting the system evolve freely in an alternating fashion~\cite{Bassler2023}.

Moreover, quantum gates can be implemented on the MAGIC platform also via microwave driving fields.
Tuning a microwave signal with frequency~$\omega_{\mathrm{MW}}$ and phase~$\phi$ close to the resonance frequency of ion~$n$ realizes the driving Hamiltonian~\cite{Wunderlich2002}
\begin{align}
    \label{eq:hdrive}
    \begin{split}
    H_{\text{drive}}(t) = \frac{\hbar \Omega}{2} \Big\{ &\etothe{i(\omega_n + \delta \omega_n - \omega_{\mathrm{MW}}) t + i \phi} \sigma_+^{(n)} \\
    &\times \exp \Big[ i \sum_l \epsilon_{ln} (a_l^\dagger \etothe{i \nu_l t} + \mathrm{h.c.}) \Big] + \mathrm{h.c.} \Big\} \,,
    \end{split}
\end{align}
where $\Omega$ is the Rabi frequency, $\delta \omega_n = \sum_l \nu_l \epsilon_{l n} / 2$ reflects a shift in the ion's resonance frequency, $\sigma_{+}^{(n)}$ is the spin raising operator acting on ion~$n$, and $\mathrm{h.c.}$ denotes the Hermitian conjugate.
This Hamiltonian is of the same form as the well-known Hamiltonian characterizing light--matter interaction in traditional laser-based trapped-ion setups and as such permits implementing the same set of quantum gates, including entangling gates like the Mølmer--Sørensen gate~\cite{Moelmer1999,Soerensen1999}, by using microwave instead of optical radiation.
In particular, single-qubit rotations can be realized by tuning the microwave frequency close to the carrier transition, $\omega_{\mathrm{MW}} \approx \omega_n + \delta \omega_n$.
In the Lamb--Dicke regime, the driving Hamiltonian in a rotating frame then becomes
\begin{align}
    \label{eq:hdrivecarrier}
    H_{\text{drive}} = \frac{\hbar\Omega}{2} \left( \etothe{i \phi} \sigma_{+}^{(n)} + \mathrm{h.c.} \right) + \frac{\hbar \Delta_n}{2} \sigma_Z^{(n)}
\end{align}
with detuning~$\Delta_n = \omega_{\mathrm{MW}} - \omega_n - \delta \omega_n$.
Rotations around an arbitrary axis in the $XY$~plane can be implemented by driving the qubits on resonance ($\Delta_n = 0$), while rotations around the $Z$~axis can be generated by a combination of $X$ and $Y$ pulses or, as is typically done, they can be realized virtually by adjusting the phase reference of the qubits~\cite{McKay2017}.

\subsubsection*{Effects and compensation of undesired local fields}

Below, we will show that one main effect of expanding the potential energy in \cref{eq:Hext} to higher order leads to additional local fields of the form $h_Z \sigma_Z$.
These terms shift the ions' resonance frequencies and result in additional contributions to unitary gates of the form $\exp(-i h_Z T \sigma_Z)$. To avoid any undesired effects, their strength~$h_Z$ should either be negligible compared to the two-body interaction $J_{ij}^{(2)}$ or these terms must actively be compensated.

To illustrate the error such a small undesired local field may generate, we consider the example of preparing the Bell state $(\ket{00} + \ket{11})/\sqrt{2}$ by applying the simple quantum circuit $\mathrm{CNOT}_{1,2} \mathrm{H}_1$ to the initial state $\ket{00}$. Here, $\mathrm{H}_1$ is a Hadamard gate acting on the first qubit and $\mathrm{CNOT}_{1,2}$ is a controlled NOT gate, with the first qubit acting as control and the second one as target. The $\mathrm{CNOT}_{1,2}$ can be further decomposed into a $ZZ$~rotation gate plus single-qubit rotations:
\begin{align} \label{eq:cnot_decomp}
   \mathrm{CNOT}_{1,2} =  \etothe{i(\pi/4)\sigma_Z^{(1)}} \etothe{i(\pi/4)\sigma_X^{(2)}} \mathrm{H}_2 \etothe{-i(\pi/4)\sigma_Z^{(1)}\sigma_Z^{(2)}} \mathrm{H}_2 \,. 
\end{align}
In Ref.~\cite{Khromova2012}, such a CNOT gate has been realized experimentally on the MAGIC platform based on always-on couplings.
Now, let us assume the $ZZ$~rotation is contaminated with small local fields, $\etothe{-i(\pi/4)\sigma_Z^{(1)}\sigma_Z^{(2)}}\rightarrow \etothe{-i(\pi/4)\sigma_Z^{(1)}\sigma_Z^{(2)}}\etothe{-ih_Z (\sigma_Z^{(1)} + \sigma_Z^{(2)})}$. Then, to still obtain the target Bell state with a fidelity higher than $\num{0.99}$, one requires $h_Z/J_{12} < \num{0.05}$. For a typical two-body interaction strength of $J_{12}^{(2)} = 2\pi \times \SI{26.5}{\hertz}$ this would correspond to $h_z / 2\pi < \SI{1.325}{\hertz}$. To put this value into context: Microwaves with frequencies that can be stabilized to the order of millihertz (mHz) through state-of-the-art technology are employed to address the qubits. This precision is much smaller than the $h_z$ required for the reliable functioning of the scheme.

In contrast, as we will see below, higher-order effects in the MAGIC scheme can lead to local fields on the same order as $J_{ij}^{(2)}$. As the above consideration illustrates, high-fidelity quantum information processing requires one to carefully compensate such undesired contributions.
In the above scenario, this may most conveniently be achieved via additional virtual $Z$ rotations~\cite{McKay2017}. This approach is generally viable for gates that commute with the error or for sufficiently small rotations. Moreover, in applications involving a microwave driving, undesired longitudinal fields may also be compensated by appropriately changing the frequency detuning of the microwave field, see \cref{eq:hdrivecarrier}.
As an alternative, systematic shifts corresponding to longitudinal fields can be cancelled via dynamical decoupling techniques. As mentioned above, e.g., in the case of encoding the qubit in magnetic-field sensitive states, dynamical decoupling is typically used in order to suppress dephasing due to fluctuations of the magnetic field, such that its use for cancelling the undesired higher-order terms does not incur relevant additional overhead.

\subsection{Example: MAGIC with \ce{^{171}_{}Yb+} ions}

\begin{figure}
    \includegraphics{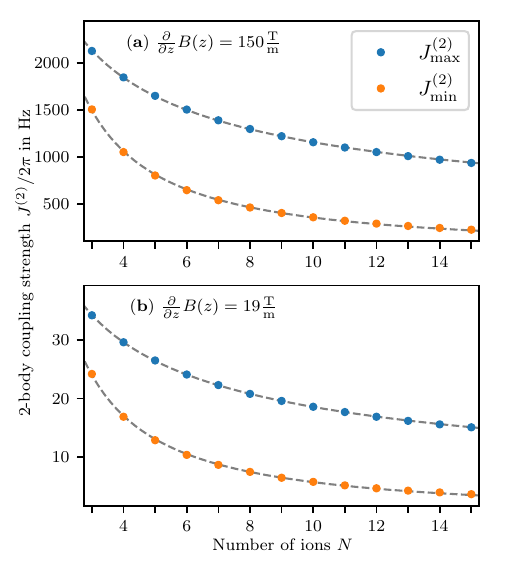}
    \caption{%
        Spin--spin coupling strengths in the MAGIC scheme, calculated as a function of the number of \ce{^{171}_{}Yb+} ions $N$ in a chain with trapping frequency $\omega_z = 2\pi \times \SI{130}{\kilo\hertz}$ and magnetic field gradient (a) $\SI{150}{\tesla\per\meter}$ and (b) $\SI{19}{\tesla\per\meter}$. Depicted are only the largest coupling strength~$J^{(2)}_{\mathrm{max}}$ (blue), occurring between two neighboring ions at the edges of the chain, as well as the smallest coupling strength~$J^{(2)}_{\mathrm{min}}$ (orange), felt by the most distant ions at opposite ends of the chain. The decrease of $J^{(2)}_{\mathrm{max}}$ follows the law $N^{-1/2}$ to good approximation, while $J^{(2)}_{\mathrm{min}}$ decreases approximately as $N^{-1.19}$. Corresponding fits are shown as grey dashed lines.}%
    \label{fig:2bodyJ_scaling}
\end{figure}

To illustrate realistically achievable spin--spin coupling strengths, we numerically calculate $J_{ij}^{(2)}$ for a chain of five \ce{^{171}_{}Yb+} ions with trapping frequency $\omega_z = 2\pi \times \SI{130}{\kilo\hertz}$ and a magnetic field gradient of $\partial_z B =\SI{19}{\tesla\per\meter}$, realized experimentally in Ref.~\cite{Piltz2016}. We consider the states $\ket{F=0, m_F = 0}$ and $\ket{F=1, m_F =  + 1}$ ($g_F \approx 1$) in the ground state hyperfine manifold as our qubit. The resonance frequency is then given by $\omega(z) \simeq  \omega_0 + \mu_{\mathrm{B}} B(z) /\hbar$~\cite{Bassler2023}.
The largest coupling strength occurs between the two outer neighboring ions at both edges of the chain: $J_{12}^{(2)} = J_{45}^{(2)} \approx 2\pi \times \SI{26.5}{\hertz}$. The weakest coupling occurs between the leftmost and outermost ion: $J_{15}^{(2)} \approx 2\pi \times \SI{12.9}{\hertz}$.
For a stronger magnetic field gradient of $\SI{150}{\tesla\per\meter}$, which is feasible with current technology~\cite{Weidt2016, SiegeleBrown2022}, the maximal (minimal) spin--spin coupling strength increases to $J_{12}^{(2)} = J_{45}^{(2)} \approx 2\pi \times \SI{1650} {\hertz}$ ($J_{15}^{(2)} \approx 2\pi \times \SI{805}{\hertz}$).

\begin{table*}[ht]
    \centering
    \begin{tabularx}{\textwidth}{YYYYYYYYYY}
    \toprule
                   && \multicolumn{4}{c}{$\partial_z B =\SI{19}{\tesla\per\meter}$} & \multicolumn{4}{c}{$\partial_z B =\SI{150}{\tesla\per\meter}$} \\
    \cmidrule(lr){3-6}\cmidrule(lr){7-10}
         ion species & N& $ \max_{ij} |J_{ij}^{(2)}|$ & $\min_{ij} |J_{ij}^{(2)}|$& $ \max_{ijk} |J_{ijk}^{(3)}|$ & $\max_{i} |h_i^z|$ & $ \max_{ij} |J_{ij}^{(2)}|$ & $\min_{ij} |J_{ij}^{(2)}|$ & $ \max_{ijk} |J_{ijk}^{(3)}|$ & $\max_{i} |h_i^z|$ \\
    \midrule
          \ce{^{171}_{}Yb+}& 5  & 26.5 & 12.9 & \SI{11.5e-6}{} & 6.2 & 1650.6 & 805.1& 0.006 & 49.2 \\
         \ce{^{171}_{}Yb+}& 10  & 18.6 & 5.8 & \SI{4.2e-6}{} & 9.6 & 1157.7 & 361.1 & 0.002 & 75.6 \\
         \ce{^{171}_{}Yb+}& 15  & 15.1  & 3.7 & \SI{2.0e-6}{} & 11.9 & 938.4  & 229.7 & 0.001 & 94.2 \\
          \ce{^{9}_{}Be+} &  5 & 125.8  & 61.3 & \SI{195e-6}{} & 22.2 & 7837.3 & 3822.9 & 0.1 & 175.1 \\
         \ce{^{9}_{}Be+}& 10  &  88.2 & 27.5 & \SI{70.8e-6}{} & 34.1 & 5496.8 & 1714.7 & 0.04 & 268.9 \\
         \ce{^{9}_{}Be+}& 15  & 71.5 & 17.5 & \SI{34.4e-6}{} & 42.5 & 4455.8 & 1090.8 & 0.02 & 335.2 \\
    \bottomrule
    \end{tabularx}
    \caption{\label{tab:coulomb_corrrections}%
    Comparison of always-on interaction strengths in the MAGIC setup with ion chains of different length ($N=5,10,15$ ions), different ion species (\ce{^{171}_{}Yb+} and \ce{^{9}_{}Be+}), and for two different magnetic field gradients ($\partial_z B =\SI{19}{\tesla\per\meter}$ and $\partial_z B =\SI{150}{\tesla\per\meter}$). As qubit states, we consider $\ket{F=0, m_F = 0}$ and $\ket{F=1, m_F =  + 1}$ for \ce{^{171}_{}Yb+}, and $\ket{F=1, m_F = 0}$ and $\ket{F=2, m_F =  + 1}$ for \ce{^{9}_{}Be+}. The table lists illustrative values of the maximal and minimal two-body coupling strength $|J_{ij}^{(2)}|$, the maximal three-body coupling strength $|J_{ijk}^{(3)}|$ due to anharmonicities in the Coulomb repulsion, as well as the largest local field correction $|h_i^z| = \sum_{jk}C_{jjk}\epsilon_{ik}$ for a ground-state cooled system (which may be amplified by higher phonon occupations, see Sec.~\ref{subsec:coulomb}). All interaction strengths are given in units of $\SI{}{\hertz} / 2 \pi$.
    }
\end{table*}

In Fig.~\ref{fig:2bodyJ_scaling}, we plot the scaling behavior of the maximal and minimal coupling strengths for different numbers of ions in the chain, for the same trapping frequency and for the same two choices of magnetic field gradients as above.
From a numerical fit (grey dashed lines), the maximal spin--spin coupling strength decreases as $N^{-0.51\pm 0.01}$~\footnote{Uncertainties in the fit parameters are computed from the estimated covariance matrix returned by the \texttt{curve\_fit} function from the SciPy library. The underlying data points are calculated up to numerical precision, i.e., without considering any errors in the physical parameters.}, matching the value $N^{-1/2}$ caused by the increasing inertia of larger ion crystals in the Cirac--Zoller gate \cite{Cirac1995} or the M{\o}lmer--S{\o}rensen gate \cite{Soerensen2000}. 

In addition, we find the minimal spin--spin interaction diminishes as $N^{-1.19\pm0.01}$.
This stronger decrease can be attributed to the large distance between the involved ions (equivalent to $N$, the length of the ion chain) in interplay with contributions from phonons beyond the center-of-mass mode, which can lead to spatially decreasing interactions \cite{Britton2012, Jurcevic2014, Richerme2014, Nevado2016, Trautmann2018}. 

In \cref{tab:coulomb_corrrections}, we compare the above values for the spin--spin coupling strength to numerical data for a chain of \ce{^{9}_{}Be+} ions (also considering different chain lengths and magnetic field gradients). For the \ce{^{9}_{}Be+} ions, we consider as the qubit states $\ket{F=1, m_F = 0}$ and $\ket{F=2, m_F =  + 1}$ from the hyperfine ground state manifold. 
The mass of the ions enters the two-body coupling strength only via $\Delta z_l$ in Eq.~\eqref{eq:J2} and we therefore have $J_{ij}^{(2)} \propto 1/m$. The numerical values in \cref{tab:coulomb_corrrections} reflect this dependence through larger coupling strengths for the \ce{^{9}_{}Be+} ions. 
Lighter ions can thus achieve faster quantum gate speeds. However, certain isotopes of heavier ion species can have other advantages, which may counterbalance the slower gate times. 
E.g., \ce{^{171}_{}Yb+} ions have a simple hyperfine level structure due to their nuclear spin of $I = 1/2$ (in contrast to $I = 3/2$ for $\ce{^{9}_{}Be+}$), which can be advantageous for cooling methods.

\section{Higher-order contributions from the potential energy}\label{sec:3body}

In the above, we have outlined the standard derivation of the effective spin--spin interaction arising in trapped-ion systems from the presence of a magnetic field gradient.
This spin--spin interaction has been achieved by expanding the external potential governing the vibrational modes and the internal energy of the pseudo-spins to second order in the displacements from their equilibrium positions. 
In what follows, we investigate sources of higher-order terms in these expansions, which may contribute to the interactions in a MAGIC trapped-ion quantum computer.

This section is devoted to higher-order terms stemming from the potential energy associated with the external degrees of freedom of the ions.

We first derive the general form of the leading corrections to the Hamiltonian, giving rise to three-body spin--spin, spin--phonon, and phonon--phonon couplings (Sec.~\ref{subsec:cubicexpansion}).
Subsequently, we address specifically the effects of cubic anharmonicities in the Coulomb repulsion (Sec.~\ref{subsec:coulomb}) as well as in the external trapping potential (Sec.~\ref{subsec:cubicexternal}).
As we show by computing their natural strength using experimentally relevant parameters, they constitute negligible perturbations or can be compensated in typical setups, though in principle a device specifically designed to enhance these higher-order contributions might leverage them for quantum information processing tasks.

\subsection{General third-order expansion of the potential}
\label{subsec:cubicexpansion}

To start with, we calculate the effect of higher-order contributions of the external potential acting on the ion crystal. In this section, we focus on the Coulomb repulsion between ions.  
These contributions will naturally occur in any trapped-ion setup using a phonon bus, though we discuss their effect here specifically for the MAGIC platform. 
We start with general derivations, which we will adapt also in the next section for anharmonicities in the trapping potential, and then specialize to the Coulomb repulsion.

Extending the expansion of the potential energy $V$ of a 1D trapped-ion chain given in Eq.~\eqref{eq:Hext} up to third order, we obtain
\begin{align}
\label{eq:coulomb_V}
\begin{split}
    V(\vec{z}) &= V(\bm{z}_0) + V^{(2)}(\bm{z})+ V^{(3)}(\bm{z}) + \bigO(\vec{z}^4) \\
    &= V(\bm{z}_0) + \sum_{i,j} \frac{A_{ij}}{2} q_i q_j + \sum_{\mathclap{i,j,k}} \frac{B_{ijk}}{6} q_i q_j q_k + \bigO(q^4) \,, 
\end{split}
\end{align}
with 
\begin{align}
    A_{ij} &= \left. \frac{\partial^2 V}{\partial q_i \partial q_j} \right|_{\bm{z}_0}, \label{eq:coulomb_A} \\
    B_{ijk} &= \left. \frac{\partial^3 V}{\partial q_i \partial q_j \partial q_k} \right|_{\bm{z}_0}. \label{eq:coulomb_B}
\end{align}
The quadratic terms form the harmonic part of the potential [see \cref{eq:uncoupled_harm}], whose eigenmodes will be used to express the higher-order terms.
As in the previous section, the matrix $A$ is diagonalized by the normal modes $\bm{Q}$, which are related to the position operators via Eq.~\eqref{eq:normal modes} and quantized through ladder operators via Eq.~\eqref{eq:quantized modes}.  
The cubic terms $B_{ijk}$ can derive either solely from a higher-order expansion of the Coulomb potential, or additionally from anharmonicities in the trapping potential.
These scenarios will be analyzed separately later.
In principle, the above expansion can be continued to any desired higher order, but the resulting effects will be subleading (see also Ref.~\cite{Loewen2003}).

We note that at this order of expansion of the external potential, the axial phonon modes couple to the transverse modes \cite{Marquet2003}. Here, we neglect this effect as we assume (i) axial and transverse modes to typically live at different energy scales and (ii) the phonon modes to be ground-state cooled. 
We discuss the contributions from transverse modes in Appendix~\ref{app:transversal}. 

Inserting the quantized normal modes, the anharmonic part of Eq.~\eqref{eq:coulomb_V} reads
\begin{multline}
    V^{(3)} = \frac{1}{6} \sum_{i,j,k} B_{ijk} \sum_{l,r,s} S_{il} S_{jr} S_{ks} \Delta z_l \Delta z_r \Delta z_s \\
    \times (a_l^\dagger + a_l)(a_r^\dagger + a_r)(a_s^\dagger + a_s) \,. \label{eq:coulomb_nm}
\end{multline}
Adding this contribution to the Hamiltonian in Eq.~\eqref{eq:HMagic} and applying the polaron transformation of Eq.~\eqref{eq:polarontrafo} yields the total Hamiltonian $\Tilde{H} = \Tilde{H}^{(2)} + \Tilde{H}^{(3)}$ of the system, given by 
\begin{multline}
\label{eq:app1_total}
    \Tilde{H} = \Tilde{H}^{(2)} + \frac{1}{6} \sum_{ijk} C_{ijk} \Big( a_i^\dagger + a_i + \sum_n \epsilon_{ni} \sigma_Z^{(n)}\Big)  \\
    \times \Big(a_j^\dagger + a_j + \sum_n \epsilon_{nj} \sigma_Z^{(n)}\Big) \Big(a_k^\dagger + a_k + \sum_n \epsilon_{nk} \sigma_Z^{(n)}\Big) \,.
 \end{multline}
Here, $\Tilde{H}^{(2)}$ is the standard MAGIC Hamiltonian in Eq.~\eqref{eq:magic_coupling}, resulting from an expansion of the potential up to second order, and we have defined
 \begin{align}
     C_{ijk} = \sum_{mnl} B_{mnl} S_{mi} S_{nj} S_{lk} \Delta z_i \Delta z_j \Delta z_k \,.
 \end{align}

Sorting the terms by their order of spin interactions, we get
\begin{align}
\label{eq:spinphonon}
    \Tilde{H} &= \Tilde{H}^{(2)} + \hbar \sum_{i<j<k} J_{ijk}^{(3)} \sigma_Z^{(i)} \sigma_Z^{(j)} \sigma_Z^{(k)} \nonumber \\
    &\hphantom{={}} + \frac{1}{2} \sum_{ijk} C_{ijk} \big(a_i^\dagger + a_i\big)\sum_{nm} \epsilon_{nj} \epsilon_{mk} \sigma_Z^{(n)} \sigma_Z^{(m)}  \nonumber\\
    &\hphantom{={}} + \frac{1}{2} \sum_{ijk} C_{ijk} \big(a_i^\dagger a_j^\dagger + a_i a_j + a_i^\dagger a_j + a_i a_j^\dagger\big) \sum_n \epsilon_{nk}\sigma_Z^{(n)} \nonumber \\
    &\hphantom{={}} + \frac{1}{6} \sum_{ijk} C_{ijk} \big( a_i^\dagger a_j^\dagger a_k^\dagger  + a_i a_j a_k   + a_i^\dagger a_j^\dagger a_k + a_i^\dagger a_j a_k^\dagger    \nonumber \\
    &\hphantom{= + \frac{1}{6} \sum_{ijk} C_{ijk} \big(}+  a_i a_j^\dagger a_k^\dagger  + a_i^\dagger a_j a_k + a_i a_j^\dagger a_k + a_i a_j a_k^\dagger \big) \,.
 \end{align}

One of the corrections to the standard MAGIC Hamiltonian $\Tilde{H}^{(2)}$ is a three-body spin interaction, given by
\begin{align}
    \label{eq:H3_Coul}
    \Tilde{H}^{(3)} = \hbar \sum_{i<j<k} J_{ijk}^{(3)} \sigma_Z^{(i)}  \sigma_Z^{(j)}  \sigma_Z^{(k)}
\end{align}
with three-body interaction strength 
\begin{align}
    \label{eq:3-body_interactionCoul}
    \hbar J_{ijk}^{(3)} &= \sum_{a,b,c,l,r,s} B_{abc} S_{al} S_{br} S_{cs} \Delta z_l \Delta z_r \Delta z_s \epsilon_{il} \epsilon_{jr} \epsilon_{ks}\nonumber \\
    & = \sum_{lrs}  C_{lrs}  \epsilon_{il} \epsilon_{jr} \epsilon_{ks} \nonumber \\
    &= \sum_{a, b, c} B_{abc} \frac{J_{ia}^{(2)}}{\partial_z \omega_a} \frac{J_{jb}^{(2)}}{\partial_z \omega_b} \frac{J_{kc}^{(2)}}{\partial_z \omega_c} \,.
\end{align}
In Eq.~\eqref{eq:H3_Coul}, we left out contributions to the local fields that arise for two coinciding indices. 
As we will see below, their strength, just like the three-body interaction, is negligible in realistic situations, but they could also be compensated easily through virtual $Z$~rotations or by adjusting the frequency detuning in the presence of a microwave driving field, see \cref{sec:discretemagicgates}. 
The other terms (second to fifth row of Eq.~\eqref{eq:spinphonon}) describe a coupling of phonon modes to two-body spin interactions and local fields, as well as the usual anharmonic phonon term appearing due to anharmonicities of the external potential \cite{Marquet2003}.

\subsection{Higher-order effects from the Coulomb interaction}\label{subsec:coulomb}

Up to here, the discussion applies to general anharmonicities in the external trapping potential. To obtain concrete order-of-magnitude estimates, we now specialize to the case of anharmonicities deriving solely from the Coulomb interactions, discussing each of the above additional contributions term by term. This complements the analysis of Ref.~\cite{Loewen2003}, where the phonon dependency was discussed within a mean field approximation assuming thermal phonon populations, with qualitatively similar conclusions.

The strength of the additional terms in Eq.~\eqref{eq:spinphonon} is determined by the cubic expansion coefficients~$B_{ijk}$. From the Coulomb repulsion, they can be evaluated as \cite{Marquet2003} 
\begin{multline}
\label{eq:B_coulomb_repulsion}
   B_{ijk} = \frac{-6 m \omega_z^2}{l} \sum_{q\neq i} \frac{u_i - u_q}{|u_i - u_q|^5} \\
   \times(\delta_{ij} - \delta_{qj})(\delta_{ik} - \delta_{qk}) \,,
\end{multline}
where $l$ is the characteristic length scale, defined via $l^3 = e^2/(4\pi \epsilon_0 m \omega_z^2)$ \cite{James1998}. We have further defined dimensionless equilibrium positions by $\bm{u} = \bm{z}_0 / l$.

\subsubsection{Three-body spin interactions}
As a concrete example, we calculate the three-body interaction strength in a one-dimensional chain of five \ce{^{171}_{}Yb+} ions in an axial harmonic trapping potential of strength $\omega_z = 2 \pi \times \SI{130}{\kilo\hertz}$ and an applied magnetic gradient of $\SI{150}{\tesla\per\meter}$. 
The numerical computation is based on a microscopic model of the ion chain, where we first find the ions' equilibrium positions and then diagonalize the phonon eigenmodes, from which we can then derive all relevant energy scales. 
In this example, we find the highest three-body interaction strength to be $J_{145}^{(3)} \approx 2 \pi \times \SI{0.006}{\hertz}$ ($J_{125}^{(3)} \approx - 2 \pi \times \SI{0.006}{\hertz}$). Compared to the smallest two-body interaction strength in the same system, $J_{15}^{(2)} \approx 2 \pi \times \SI{805}{\hertz}$, the three-body spin interactions introduced by anharmonicities in the Coulomb repulsion are negligible.

Analogous numerical values for longer ion chains, as well as for a different ion species (\ce{^{9}_{}Be+}), are shown in \cref{tab:coulomb_corrrections}. Even though the maximal three-body interaction strength is an order of magnitude larger for the lighter ions, it is still negligible compared to the---also increased---two-body interactions.

\subsubsection{Phonon-nonconserving two-body spin interactions}
The term in the second row of Eq.~\eqref{eq:spinphonon} gives corrections of the form $\sim (a_i^\dagger + a_i)\sigma_Z^{(n)} \sigma_Z^{(m)}$ to the MAGIC two-body spin couplings. Their strength scales as $C_{ijk}\epsilon_{nj} \epsilon_{mk}$, and is thus one order in the Lamb--Dicke parameter $\epsilon$ stronger than $J_{ijk}^{(3)}$. 
Formally, they are of the same order in the Lamb--Dicke parameter as the two-body interaction $J_{ij}^{(2)}$. 
However, the terms $\sim (a_i^\dagger + a_i)\sigma_Z^{(n)} \sigma_Z^{(m)}$ are suppressed by another factor $C_{ijk}$; even more, in contrast to $J_{ij}^{(2)}$, they do not preserve phonon number, and can thus safely be neglected within a rotating wave approximation: In the interaction picture given by the diagonal phonon Hamiltonian $\hbar\sum_n \nu_n a_n^\dagger a_n$, the creation and annihilation operators gain a time dependence $a_i^\dagger \to e^{i\nu_i t} a_i^\dagger$. They thus rotate on a time scale significantly faster than $C_{ijk}\epsilon_{nj} \epsilon_{mk}$. 
For instance, in the above example of a chain of five \ce{^{171}_{}Yb+} ions, the strength of the terms in the second row of Eq.~\eqref{eq:spinphonon} is $|C_{ijk}\epsilon_{nj}\epsilon_{mk}| \lesssim \SI{e-1}{\hertz}$. 
This value corresponds to interaction scales that are weaker than typical coherence times in trapped-ion hardware \cite{Pogorelov2021, Schindler2013}. Moreover, it is orders of magnitude smaller than even the smallest axial phonon frequency, $\omega_\mathrm{COM} = \omega_z = 2 \pi \times \SI{130}{\kilo\hertz}$, such that this contribution can be neglected in a rotating wave approximation.

\subsubsection{Spin-dependent phonon pair production and annihilation}
The term in the third row of Eq.~\eqref{eq:spinphonon} has parts with different physical interpretation, which need to be discussed separately.  
In the same interaction picture as above, the terms creating (and annihilating) phonon pairs $a_i^\dagger a_j^\dagger$ (and $a_i a_j$) rotate with a frequency $\nu_i+\nu_j$, which is much larger than the associated energy scale $C_{ijk}\epsilon_{nk}$. 
Again, for the above example with five ions, we get corrections on the order of $|C_{ijk}\epsilon_{nk}| \lesssim \SI{e0}{\hertz}$.
As the phonon frequencies are orders of magnitude larger, these terms can again safely be neglected within a rotating wave approximation. 

\subsubsection{Spin-dependent phonon hopping}
In contrast, the number-preserving terms of the form $a_i^\dagger a_j$ rotate with the frequency $\left|\nu_i - \nu_j\right|$. For axial phonons and small systems, the difference of phonon mode frequencies is large (e.g., for a chain of five ions in a trapping potential of $\omega_z = 2 \pi \times \SI{130}{\kilo\hertz}$, phonon modes are separated by at least about $2 \pi \times \SI{80}{\kilo\hertz}$).
Terms with $i\neq j$ thus still rotate rapidly as compared to the time scale $C_{ijk}\epsilon_{nk}$ and can be neglected. 

\subsubsection{Phonon-occupation-dependent longitudinal field}
\begin{figure}
    \includegraphics{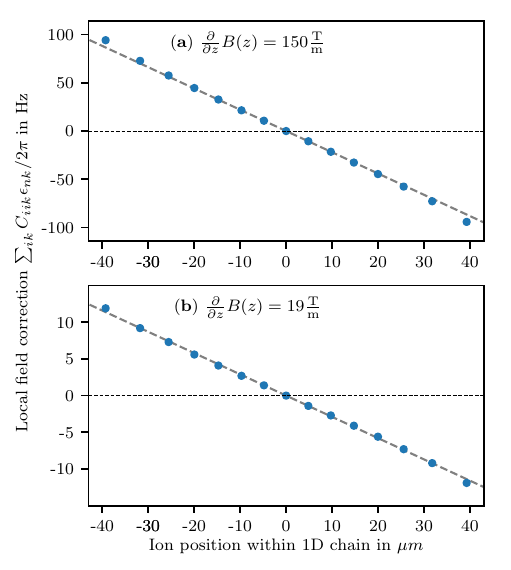}
    \caption{
    Corrections to the local field acting on each ion~$n$ (numbered from left to right) due to the anharmonicity of the Coulomb repulsion, $\sum_{ik} C_{iik} \epsilon_{nk} \sigma_Z^{(n)}$ [third line in Eq.~\eqref{eq:spinphonon}]. 
    Values computed for a chain of 15 \ce{^{171}_{}Yb+} ions and plotted over their spatial positions, with a trapping frequency of $\omega_z = 2\pi \times \SI{130}{\kilo\hertz}$ and magnetic field gradients of (a) $\SI{150}{\tesla\per\meter}$ and (b) $\SI{19}{\tesla\per\meter}$, assuming ground-state-cooled phonons.
    These corrections are highly inhomogeneous and can acquire significant values comparable to spin--spin couplings (see Fig.~\ref{fig:2bodyJ_scaling}). The grey dashed line is a linear fit through the central three ions.
    }
    \label{fig:local_corrections}
\end{figure}
We are thus left with terms where $i= j$, which give a correction to the local fields of strength $\sim~\sum_{ik} C_{iik} \epsilon_{nk} {(2n_i + 1)}$, where $n_i = a_i^\dagger a_i$. 
Again, in the example of a chain of five \ce{^{171}_{}Yb+} ions with $\omega_z = 2 \pi \times \SI{130}{\kilo\hertz}$ and a magnetic gradient of $\SI{150}{\tesla\per\meter}$, and assuming zero phonon number excitations ($n_i = 0$), we get corrections on the order of $\SI{e1}{\hertz}$.
These are largest for the outermost ions at the edges of the chain ($n = 1$ and $n = 5$), amounting to $\sum_{ik}C_{iik}\epsilon_{nk} \approx \pm  2 \pi \times \SI{49.2}{\hertz}$.
For a magnetic gradient of $\SI{19}{\tesla\per\meter}$, as is used in current experiments \cite{Piltz2016}, this magnitude drops to $2 \pi \times \SI{6.2}{\hertz}$. For non-zero phonon occupation numbers, these additional longitudinal fields get further enhanced by factors of $(2n_i + 1)$ in each term of the summation. The local field corrections are comparable to or even larger (for non-zero phonon occupation numbers)  than the two-body coupling strengths $J^{(2)}/2\pi \approx \SIrange{13}{27}{\hertz}$. They may generate an additional contribution to the two-qubit gates when implementing them as explained in Sec.~\ref{sec:magic} (the magnitude of such gate errors has also been previously discussed in Ref.~\cite{Loewen2003}).

In Fig.~\ref{fig:local_corrections}, we show the behavior of the local field corrections across a chain of 15 ions for magnetic field gradients of (a) $\SI{150}{\tesla\per\meter}$ and (b) $\SI{19}{\tesla\per\meter}$. The corrections are largest towards the edges of the chain (equal magnitudes with opposite signs) and vanish for the ion at the center. The magnitude of the correction scales approximately linearly with the distance to the center of the chain, with larger deviations toward the edges. This is visualized by the grey dashed line, which is a linear fit through the centermost three ions. 
Comparing the largest corrections at the edges of the chain in Fig.~\ref{fig:local_corrections} (b) ($\approx 2\pi \times \SI{11.9}{\hertz}$) to the spin--spin coupling strengths shown in Fig.~\ref{fig:2bodyJ_scaling} (b) ($J_{\mathrm{max}}^{(2)}/2\pi \approx \SI{15.1}{\hertz}$), we find them to be already of the same order of magnitude as the always-on spin--spin couplings for this system size.

\begin{figure}
    \includegraphics{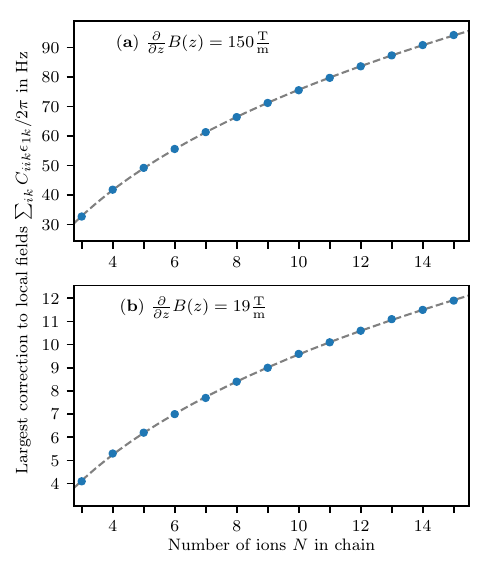}
    \caption{
    Scaling of the largest correction $\sum_{ik} C_{iik} \epsilon_{1k} \sigma_Z^{(1)}$ (on the leftmost edge) to the local fields with increasing number of \ce{^{171}_{}Yb+} ions in the chain. Calculations were done for a vanishing phonon occupation, a trapping frequency of $\omega_z = 2\pi \times \SI{130}{\kilo\hertz}$, and magnetic gradients of (a) $\SI{150}{\frac{\mathrm{T}}{\mathrm{m}}}$ and (b) $\SI{19}{\frac{\mathrm{T}}{\mathrm{m}}}$. The absolute strength increases with chain length, rendering this correction significant in particular when scaling to larger ion chains. Numerical fits (grey dashed lines) show a scaling $\propto N^{0.18} \log (1.38 N)$.    
    }
    \label{fig:scaling_analysis}
\end{figure}

In Fig.~\ref{fig:scaling_analysis}, we plot the scaling of these largest corrections at the edges of the chain with system size, again for the same values of the magnetic field gradient.
We find that the local field corrections increase with the length of the chain. 
In contrast, as illustrated above in Fig.~\ref{fig:2bodyJ_scaling}, the spin--spin coupling strengths decrease. 
Thus, these local corrections become more relevant as the system size increases. 
Phenomenologically, we find the scaling behaviour to be best described by a power law with logarithmic corrections of the form $N^a \log (bN)$, with fitting parameters $a = 0.18 \pm 0.01$, $b = 1.38 \pm 0.04$ for $\SI{150}{\tesla\per\meter}$ and $a = 0.18 \pm 0.01$, $b = 1.35 \pm 0.07$ for $\SI{19}{\tesla\per\meter}$.

For comparison, \cref{tab:coulomb_corrrections} shows the maximal (at the edges of the chain) local field corrections also for chains of lighter \ce{^{9}_{}Be+} ions. For a gradient of $\SI{19}{\tesla\per\meter}$, the corrections are again on the same order of magnitude as the always-on two-body coupling strengths. The ratio improves with a larger magnetic field gradient, but worsens with increased chain length.

The presence of an undesired local field could constitute an issue for the precise functioning of two-qubit gates or for the correct definition of cost functions in quantum optimization \cite{Hauke2020}. 
Fortunately, since the associated terms are proportional to $\sigma_Z^{(n)}$, even for nonzero mean phonon excitation numbers they can be offset by adjusting the detuning of a microwave driving field from the qubit resonances or performing virtual Z gates \cite{McKay2017} (see Sec.~\ref{sec:discretemagicgates}). 
In typical trapped-ion devices, such a compensation is often routine, even for terms many orders of magnitude stronger than in this case. For instance, in a M{\o}lmer--S{\o}rensen setup on the axial modes of \ce{^{40}_{}Ca+} ions, undesired off-resonant excitation of dipole transitions can induce inhomogeneous light shifts on the order of $\SIrange{2}{3}{\kilo\hertz}$, which need to be calibrated and compensated~\cite{Jurcevic2014}. 
In addition, already simple dynamical-decoupling techniques can remove undesired terms $\sim \sigma_Z^{i}$~\cite{Viola1998, Loewen2003}.

While the mean phonon number contribution to these local fields can easily be compensated, phonon number fluctuations remain problematic as they lead to fluctuating shifts. This underscores the critical importance of cooling techniques, such as sympathetic cooling \cite{Rohde2001, Sriarunothai2017, Sosnova2021}, to minimize these fluctuations and ensure precise gate operations.

\subsubsection{Phonon creation, annihilation, and conversion terms}
Finally, the two last lines of Eq.~\eqref{eq:spinphonon} contain terms that describe the generation/annihilation of triples of phonons as well as two-to-one phonon conversion, which have been discussed before \cite{Marquet2003}. The former terms can be neglected since they rotate with at least three times the trap frequency, which is much larger than the interaction strength~$C_{ijk}$. The two-to-one conversion terms, however, require careful attention. As long as all modes are ground-state cooled, these terms vanish. 
They can also be suppressed by designing the trap parameters such that the phonon frequencies do not conspire to render such terms resonant. 
For the above setup with five ions and a magnetic field gradient of $\SI{150}{\tesla\per\meter}$, the smallest frequency difference $\left| \nu_i + \nu_j - \nu_k \right|$ is about $\SI{27}{\kilo\hertz}$. In this specific case, these terms can thus safely be neglected even if the chain is not fully cooled into the motional ground state. 
Nevertheless, these caveats need to be kept in mind especially when scaling up to large ion chains, where ground-state cooling becomes more challenging and resonances can appear more easily. 

To summarize this part, among the various corrections to the MAGIC trapped-ion scheme that derive from a third-order expansion of the Coulomb repulsion, only phonon-number-dependent longitudinal fields as well as two-to-one phonon conversions can become sizable. The effect of the former
can be compensated via additional virtual $Z$~rotations in the case of discrete quantum gates~\cite{McKay2017} or, in applications involving a microwave driving, by adjusting the frequency detuning,
while the latter can be avoided by ground-state cooling of the phonon modes.

\subsection{Cubic anharmonicities of the external trapping potential}\label{subsec:cubicexternal}

In addition to the intrinsic anharmonicities given by the Coulomb potential, there could also be anharmonicities stemming from the external trapping potential.
While such effects are often undesired for quantum logic, anharmonicities can nowadays also be designed and exploited in a controlled way~\cite{Schmied2008, Johanning2016}.
Such anharmonic potentials can change the equilibrium positions of the ions and therefore modify the two-body interactions induced by a magnetic field gradient.
This effect has been investigated in Ref.~\cite{Zippilli2014}.
Here, we discuss the role of higher-order terms generated by anharmonicities in the external potential.

To this end, we consider a cubic contribution to the potential energy of the form
\begin{align}
\label{eq:local_cubic_potential}
    V^{(3)}(\vec{z}) &=  \sum_n \frac{\alpha_n}{6} (z_n - z_{n,0})^3 \,,
\end{align}
where $\alpha_n$ describes the strength of the cubic anharmonicity felt by ion~$n$.
Such terms in the potential energy generate additional three-body spin--spin, spin--phonon, and phonon--phonon interactions, for which the general considerations discussed in Sec.~\ref{subsec:coulomb} apply.
Here, we investigate the resulting three-body spin interaction in more detail.
 
Inserting the expansion coefficients $B_{ijk}=\delta_{ij}\delta_{jk} {\alpha_i}$ into \cref{eq:3-body_interactionCoul}, the three-spin coupling coefficients specialize to
\begin{equation}
\label{eq:J3_trap}
    \hbar J_{ijk}^{(3)} = \sum_{n} \frac{\alpha_n}{(\partial_z \omega_n)^3} J_{in}^{(2)} J_{jn}^{(2)}J_{kn}^{(2)} \,.
\end{equation}

In order to obtain a simple order-of-magnitude estimate for their typical strength, we first estimate the typical strength of the two-body spin interaction in \cref{eq:J2}.
For this purpose, we consider only the contribution from the center-of-mass (cm) mode, whose frequency coincides with the trap frequency, $\nu_{\mathrm{cm}} = \omega_z$, and whose matrix elements $S_{i,\mathrm{cm}}$ are given by $1 / \sqrt{N}$.
Assuming that the gradient of the resonance frequency is the same for all ions ($\partial_z \omega_i = \partial_z \omega$), we have $J^{(2)} \simeq \omega_z \epsilon^2$ 
with the effective Lamb--Dicke parameter $\epsilon = \partial_z \omega \Delta z / \omega_z \sqrt{N}$ and oscillator length $\Delta z = \sqrt{\hbar / 2 m \omega_z}$.
Combining this expression with \cref{eq:J3_trap}, we can estimate the typical strength of the three-body coupling as
\begin{align}
\label{eq:J3_trap_estimate}
    \hbar J^{(3)}
    \simeq \frac{1}{\sqrt{N}} \frac{\alpha \Delta z ^3}{\hbar \omega_z} \mathinner{\epsilon} \hbar J^{(2)}
    = \frac{1}{\sqrt{N}} \alpha \Delta z^3 \epsilon^3 \,,
\end{align}
where we have made the additional simplifying assumption that the anharmonicity is the same at each ion position ($\alpha_n = \alpha$).

One may be tempted to engineer cubic terms in the potential similar to \cref{eq:local_cubic_potential} with the goal of harnessing the resulting three-body spin interaction for quantum information processing tasks.
Unfortunately, as we argue in what follows, this endeavor is unlikely to be advantageous in practice.

To this end, we note that cubic contributions to the potential energy may severely affect the stability of the trap.
For instance, an external potential of the form $V_{\mathrm{trap}}(\vec{z}) = \sum_n m \omega_z^2 z_n^2 / 2 + \sum_n \alpha z_n^3 / 6$ with global anharmonicity~$\alpha$ gives rise to terms as in \cref{eq:local_cubic_potential} with $\alpha_n = \alpha$.
However, such a potential also shifts the equilibrium positions of the ions.
Thus, in order to prevent ions from being ejected from the trap, the anharmonicity should be a somewhat small perturbation on top of the harmonic part, such that the potential still features a local minimum that is sufficiently deep and wide to trap the ions.
That is, the energy scale $\alpha \Delta z^3$ should be well below the harmonic trapping energy $\hbar \omega_z$, requiring $\alpha \Delta z^3 / \hbar \omega_z \ll 1$.
On the other hand, the three-body interaction should be sufficiently strong in order to be of practical use.
For example, a native three-qubit gate should outperform the equivalent gate synthesized from a combination of two-qubit gates.
Thus, it is desirable that $J^{(3)} = \lambda J^{(2)}$ with $\lambda$ as large as possible (though typically well below unity).
The estimate in \cref{eq:J3_trap_estimate} can then be expressed as $\alpha \Delta z^3 / \hbar \omega_z = \lambda \sqrt{N} / \epsilon$.
As noted before, stability of the trapping potential requires this quantity to be sufficiently small, enforcing the hierarchy $\lambda = J^{(3)} / J^{(2)} \ll \epsilon / \sqrt{N}$.
This means that the three-qubit interaction is bound to be weaker than the two-qubit interaction by a factor well below $\epsilon / \sqrt{N}$, which in realistic setups amounts to a suppression by at least one, if not several, orders of magnitude.

To circumvent the stability issues described above, one may attempt to engineer the terms in \cref{eq:local_cubic_potential} \emph{locally}, i.e., without affecting the equilibrium positions of the ions, as may be possible, e.g., in segmented ion traps~\cite{Schmied2008}.
However, from a practical point of view, the electrodes creating such local potentials cannot be placed arbitrarily close to the ions, but require a certain minimum distance, which is typically large compared to the spacing between ions~\cite{Schmied2008}.
Consequently, it becomes difficult to create localized features varying on the scale of the oscillator length~$\Delta z$, which is much smaller than the distance between ions.
For example, it appears challenging to engineer localized external potentials stronger than the intrinsic Coulomb potential created by the ions themselves.
Even if the strength of the local cubic potential was comparable to the characteristic energy scale of the anharmonic part of the Coulomb repulsion in \cref{eq:B_coulomb_repulsion}, $\alpha \Delta z^3 = 6 m \omega_z^2 \Delta z^3  / l \approx \num{0.004} \, \hbar \omega_z$, the maximum three-body coupling strength would amount to only $J_{\mathrm{max}}^{(3)} / 2 \pi \approx \SI{0.3}{\hertz}$ in a chain of $N = 5$ ions with trapping frequency $\omega_z / 2 \pi = \SI{130}{\kilo\hertz}$ and magnetic field gradient $\partial_z B = \SI{150}{\tesla\per\meter}$, which is too low for typical applications.
For comparison, a potential with a \emph{global} cubic anharmonicity of the same strength would induce equally weak three-body interactions, but would not be able to trap the ions for the above parameters in its local minimum.

To sum up, while cubic anharmonicities can in principle be a route to engineering three-body interactions in the MAGIC setup, the comparatively slow three-qubit interaction rate and the substantial experimental overhead of realizing the required cubic anharmonicities in a stable way limit the practical use of this scheme.
After all, our discussion reveals that moderate cubic anharmonicities typically have a negligible parasitic impact on the standard MAGIC setup.

\section{Higher-order contributions from the magnetic field curvature}\label{subsec:higherorderB}

While anharmonicities in the external potential of the ion crystal appear in any trapped-ion platform, a higher-order contribution specific to the MAGIC setup originates from higher-order spatial derivatives of the magnetic field. To evaluate their effect, we expand the magnetic-field-dependent resonance frequency of the ions [see Eq.~\eqref{eq:resonancefrequency_gradient}] up to second order,  
\begin{align}
\begin{split}
\label{eq:quadratic_resonance_frequency_expansion}
    \omega(z_n) &= \omega_n + \partial_z \omega(z_n)\Big\rvert_{z_{n,0}} (z_n - z_{n,0}) \\ &\hphantom{={}} + \frac{1}{2}\partial^2_z \omega (z_n) \Big\rvert_{z_{n,0}} (z_n - z_{n,0})^2 + \bigO \left( z_n^3 \right) \,.
\end{split}
\end{align}

The Hamiltonian $H_\mathrm{int}$ in Eq.~\eqref{eq:Hint}, describing the internal energy of the ion chain, then becomes
\begin{align}
    H_\mathrm{int} = - \frac{\hbar}{2} \sum_n \Big( \omega_n + \partial_z \omega_n q_n + \frac{1}{2} \partial^2_z \omega_n q_n^2 \Big) \sigma_Z^{(n)} \,. \label{eq:H_I_1}
\end{align}

Inserting again the quantized normal modes, the internal energy Hamiltonian reads
\begin{multline}
\label{eq:curvature_full}
    H_\mathrm{int} = - \frac{\hbar}{2} \sum_n \omega_n \sigma^{(n)} - \frac{\hbar}{2} \sum_{nl} \partial_z \omega_n S_{nl} \Delta z_l \big(a_l^\dagger + a_l\big)  \sigma_Z^{(n)} \\
    - \frac{\hbar}{4} \sum_{nlr} \partial^2_z \omega_n   S_{nl} S_{nr} \Delta z_l \Delta z_r \big(a_l^\dagger + a_l\big) \big(a_r^\dagger + a_r\big) \sigma_Z^{(n)} \,.
\end{multline}

Performing, as before, the polaron transformation to decouple phonons and spins in the first order terms, Eq.~\eqref{eq:polarontrafo}, we obtain the total Hamiltonian 
$\Tilde{H}=\Tilde{H}^{(2)}+\Tilde{H}^{(3)}$ 
with 
\begin{align}
\begin{split}
    \Tilde{H}^{(3)} &=  - \frac{\hbar}{4} \sum_{nlr} \partial_z^2 \omega_n S_{nl} S_{nr} \Delta z_l \Delta z_r \sigma_Z^{(n)} \\
    &\times \Big( a_l^\dagger + a_l + \sum_i \epsilon_{il} \sigma_Z^{(i)} \Big) \Big( a_r^\dagger + a_r + \sum_j \epsilon_{jr} \sigma_Z^{(j)} \Big) \,.
\end{split}
\end{align}
Again, $\Tilde{H}^{(2)}$ is the standard MAGIC Hamiltonian, while $\Tilde{H}^{(3)}$ are the corrections due to the expansion of the magnetic field distribution up to third order.
We can sort all terms by their power in the spin operators and write
\begin{align}
\begin{split}
    \Tilde{H} &= \Tilde{H}^{(2)} - \frac{\hbar}{4} \sum_{ijk} J_{ijk}^{(3)} \sigma_Z^{(i)} \sigma_Z^{(j)} \sigma_Z^{(k)} \\
    &\hphantom{={}}- \frac{\hbar}{2} \sum_{ijlr} \Tilde{C}_{ilr} \epsilon_{jr} \big(a_l^\dagger + a_l\big) \sigma_Z^{(i)}\sigma_Z^{(j)} \\
    &\hphantom{={}}- \frac{\hbar}{4} \sum_{ilr} \Tilde{C}_{ilr} \big(a_l^\dagger a_r^\dagger + a_l^\dagger a_r + a_l a_r^\dagger + a_l a_r\big) \sigma_Z^{(i)} \,,
\end{split}
\end{align}
where we defined 
\begin{align}
    \Tilde{C}_{ilr} = \partial_z^2 \omega_i S_{il} S_{ir} \Delta z_l \Delta z_r \,.
\end{align}

Analogously to the discussion of Sec.~\ref{subsec:coulomb}, the terms in the second row give corrections to the two-body spin interactions of the form $(a_l^\dagger + a_l) \sigma_Z^{(i)}\sigma_Z^{(j)}$. Again, these do not preserve phonon numbers and can be neglected within a rotating wave approximation. The terms in the third row only give corrections to local fields of strength $\sim \Tilde{C}_{irr}(2n_r + 1)$ with $n_r = a_r^\dagger a_r$ (neglecting phonon-number-non-preserving terms as before), which can be offset by microwave detunings or neglected for small magnetic field curvatures ($\Tilde{C}_{irr} \propto \partial_z^2 \omega_i$).
Due to the different origin of the higher-order terms as compared to Sec.~\ref{subsec:coulomb}, there are no cubic phonon terms in this expansion. 

The remaining effective three-spin interaction is

\begin{align}
    \Tilde{H}^{(3)} &= - \frac{\hbar}{4} \sum_{ijk}  J_{ijk}^{(3)}\sigma_Z^{(i)}\sigma_Z^{(j)} \sigma_Z^{(k)} \label{eq:H_I_2}
\end{align}
with the three-body coupling strength
\begin{align}
    J_{ijk}^{(3)} &= \sum_{lr} \partial^2_z \omega_i   S_{il} S_{ir} \Delta z_l \Delta z_r \sum_{jk} \epsilon_{jl} \epsilon_{kr} \,.
\end{align}

The form of this term differs from the ones in the previous section [Eqs.~\eqref{eq:3-body_interactionCoul} and~\eqref{eq:J3_trap}], since it did not derive from a contribution $\propto x^3$, but rather $\propto x^2\sigma^z$. 

By defining $\gamma_n= \partial^2_z \omega_n / (\partial_z \omega_n)^2$, i.e., the ratio between the magnetic field curvature and the square of the gradient, and using the spin--spin coupling strength given in Eq.~\eqref{eq:J2}, one can write this interaction even more compactly as 
\begin{align}
J_{nij}^{(3)} = \gamma_n J_{ni}^{(2)} J_{nj}^{(2)} \,.
\end{align}
Notably, $J_{nij}^{(3)} = J_{nji}^{(3)}$, but in general $J_{nji}^{(3)} \ne J_{jni}^{(3)}$.  
One can symmetrize the 3-body coupling strength to   
\begin{align}
    \Tilde{J}_{ijk}^{(3)} &= \gamma_i J_{ij}^{(2)} J_{ik}^{(2)} + \gamma_j J_{ji}^{(2)} J_{jk}^{(2)} + \gamma_k J_{ki}^{(2)} J_{kj}^{(2)} \,, 
\end{align}
which assumes a form similar to that obtained in Ref.~\cite{Bermudez2009} for a trapped-ion system driven close to the second sideband. 

In analogy to the procedure for anharmonic external potentials in Sec.~\ref{subsec:cubicexternal}, the typical strength of the three-body coupling induced by the magnetic field curvature can be estimated as
\begin{equation}
\label{eq:J3_curvature_estimate}
    J^{(3)} = \frac{1}{\sqrt{N}} \frac{\partial_z^2 \omega \Delta z}{\partial_z \omega} \mathinner{\epsilon} J^{(2)} = \frac{1}{N} \partial_z^2 \omega \Delta z^2 \epsilon^2 \,.
\end{equation}
Remarkably, this quantity scales quadratically with the effective Lamb--Dicke parameter~$\epsilon$ and thus with the magnetic field gradient, like the two-body coupling $J^{(2)}$.

For a potential application of this \enquote{magnetic curvature induced three-body coupling} in quantum information processing, the quadratic variation of the local resonance frequency $\partial_z^2 \omega$ over length scales on the order $\Delta z$ should be sufficiently large.
However, similarly as for the local cubic potentials discussed in \cref{subsec:cubicexternal}, in practice it is very hard to engineer magnetic field distributions with sufficiently sharp localized features.

Instead, we consider in what follows a \emph{global} magnetic field curvature of the form $\vec{B}(z) = \vec{B}_0 + (\mathinner{\partial_z B} z + \mathinner{\partial_z^2 B} z^2 / 2) \vec{\hat{e}}_z + \bigO(z^3)$.
Unlike higher-order terms in the external potential, higher-order contributions to the magnetic field do not alter the equilibrium positions of the ions as they act on the internal degrees of freedom, but they can lead to a position-dependent magnetic-field gradient~$\partial_z B(z_n)$.
As an example, we consider a magnetic field curvature of $\partial^2_z B \simeq \SI{e3}{\tesla\per\meter\squared}$, which can be reached using suitably designed Halbach magnets~\cite{Bluemler2023}.
If the ion chain is centered around the origin, the variation of the magnetic field gradient is negligible, $\partial B(z_n) / \partial z = \partial_z B + z_n \partial_z^2 B \approx \partial_z B$, since the spacing between ions is on the order of \SI{10}{\micro\meter}.
In addition, for five\ce{^{171}_{}Yb+} ions with a magnetic field gradient of $\partial_z B = \SI{150}{\tesla\per\meter}$ and a trap frequency of $\omega_z / 2\pi = \SI{130}{\kilo\hertz}$, we obtain three-body coupling strengths of only $J^{(3)} \sim 2\pi \times \SI{e-5}{\hertz}$, which is orders of magnitude too low for practical use.
Nonetheless, this example underlines that, as long as the magnetic field gradient does not significantly vary on length scales comparable to the oscillator length~$\Delta z$, parasitic effects from nonlinear gradients can safely be neglected in the standard MAGIC scheme.

\section{Comparison to existing schemes for generating higher-order interactions in trapped ions}\label{subsec:3body}

As we have seen in the above, higher-order terms in the MAGIC setup naturally lead to three-body interactions and spin--phonon couplings. We have found that the three-body interaction strength is negligible compared to the two-body interactions for realistic experimental setups. For instance, the corrections arising due to inherent anharmonicities in the Coulomb repulsion only produce a three-body interaction strength on the order of $\SI{e-2}{\hertz}$ for typical parameters (see \cref{subsec:coulomb}).

However, since such types of interactions are key ingredients in applications ranging from quantum simulation of topological and Peierls phase transitions \cite{GonzalezCuadra2018, GonzalezCuadra2019} and lattice gauge theories \cite{Yang2016,Andrade2022,Mildenberger2022} to improved schemes of quantum annealing \cite{Nagies2025a,Lechner2015}, one may wonder if there are alternative feasible ways to engineer such interactions and how the achievable interaction strengths compare to the ones we have discussed above. 

Indeed, various proposals have been designed for this purpose specifically with trapped-ion setups in mind. 
One rather generally applicable approach relies on obtaining three-body interactions perturbatively from two-body terms, e.g., 
via an energy penalty~\cite{Hauke2013}
or via a high-frequency expansion based on Floquet engineering 
\cite{Decker2020}. 
The 3-body interaction in Ref.~\cite{Hauke2013} arises from perturbation theory, combining an effective magnetic field acting on the pseudo-spins, $\sim K$, and a M\o lmer--S\o rensen interaction $\sim J$. These interactions are further suppressed by another M\o lmer--S\o rensen interaction $\sim V$, generated by an additional pair of beams, which is assumed to be the largest energy scale. Within perturbation theory, the 3-body strength is therefore of order $KJ/V$. The numerics in Ref.~\cite{Hauke2013} show that values of around $J=K=0.3 \, V$ are sufficient to enter the relevant perturbative regime. Assuming that $V$ corresponds to reasonably achievable interaction strengths from M\o lmer--S\o rensen gates on the order of $\sim 2\pi \times\SI{1}{\kilo\hertz}$, this results in a three-body interaction strength on the order of $0.3^2 \, V\approx 2\pi \times\SI{0.1}{\kilo\hertz}$.
As a feature of the approach in Ref.~\cite{Hauke2013}, in the effective Hamiltonian the two-body interactions play no role any more. 

Other approaches are based on continuously and simultaneously addressing the first and second sideband \cite{Bermudez2009,Andrade2022}, which results in the simultaneous presence of terms that are formally similar to  the terms $q_n \sigma_Z^{(n)}$ and $q_n^2 \sigma_Z^{(n)}$ that appear in Eq.~\eqref{eq:H_I_1}. 
In this approach, the ability to use large Rabi frequencies on the first and second sideband enables large three-body interactions whose strength can be tuned to match that of the two-body interactions. The authors of Ref.~\cite{Bermudez2009} discuss 
how the dipole force can be adapted, assuming reasonable 
trap frequencies $\sim \SI{10}{\mega\hertz}$, 
laser detunings $\sim \SI{1.25}{\mega\hertz}$, and 
stiffness parameter $\sim 0.05$, 
to reach a three-body interaction strength of about $2 \pi \times \SI{0.1}{\kilo\hertz}$, and, by relaxing trap frequencies or designing specific microtraps, even up to about $2 \pi \times \SI{1.6}{\kilo\hertz}$ \cite{Bermudez2009}. 

A further promising proposal has recently been put forward based on a combination of discrete, spin-dependent displacement and squeezing pulses \cite{Katz2022}. In a chain of four \ce{^{171}_{}Yb+} ions, the authors estimate the gate time of a fully entangling three-qubit gate, $U = \exp \big(-i\frac{\pi}{4} \sigma_Z^{(i)} \sigma_Z^{(j)} \sigma_Z^{(k)}\big)$, to be $T \approx \SI{130}{\micro\second}$ for feasible setups with laser-based control, corresponding to a three-body coupling strength on the order of $2\pi \times \SI{1}{\kilo\hertz}$.

The above three proposals rely only on ingredients already demonstrated in trapped ions, though their combination to obtain a three-body interaction has thus far not been demonstrated in experiment.

Finally, a standard way that is always possible given a universal gate set is to engineer the unitary three-qubit interactions through a combination of two-qubit and single-qubit gates. For example, a term of the form $\sigma_Z^{(i)}\sigma_Z^{(j)} \sigma_Z^{(n)}$ can be realized using four M\o lmer--S\o rensen gates, see, e.g., Ref.~\cite{Nagies2025a}.
Neglecting the gate time of single-qubit operations, one can associate an effective three-body strength to this gate sequence of order $J^{(3)}=J^{(2)}/4$, which is a rather sizable interaction strength. As the unitary three-body gate is generated exactly in this case, there are no additional two-body interactions. 

In addition to three-spin interactions, there are also interesting applications involving higher-order products of spin and phonon operators, such as terms of the form $a_i a_j \sigma_Z^{(n)}$ or $a_i \sigma_Z^{(n)} \sigma_Z^{(m)}$. Above, we have seen that the higher-order expansion can generate such terms as small corrections.   
In Ref.~\cite{Yang2016}, it has been shown how these terms can be used to quantum simulate a lattice version of quantum electrodynamics. In that work, it was found that near-resonant addressing of first and second sidebands may generate interaction strengths up to $ 2\pi\times \SI{120}{\hertz}$.

\section{Conclusions}\label{sec:conclusion}

To summarize, in this article we have presented a thorough analysis of how higher-order terms in the external potential and the magnetic field take effect in the MAGIC scheme. 
While most of our discussion was phrased in terms of shifts of spin interaction strengths and local fields, the results can be straightforwardly applied to discrete gates by considering finite gate times and hiding qubits not involved in the gate in other internal states that are insensitive to the magnetic field.
We have derived the effective interactions that result on top of the desired two-body interactions after a polaron transformation on the axial phonon modes. 
There are two generic main contributions: phonon--spin couplings and three-body spin--spin interaction terms. 
As we have shown through our analysis, in typical situations such higher-order terms can safely be neglected as compared to the two-body interactions, and in particular also as compared to other error sources such as deriving from fluctuations or drifts in trap frequency, microwave or laser intensity, timing precisions, heating of phonon modes, spontaneous emission, etc \cite{Barthel2023}. 
Since such interaction terms may also be desired for certain applications, e.g., quantum simulation of strongly-correlated systems or enhanced quantum-optimization schemes \cite{Nagies2025a}, we have put these values in context by giving a brief survey on existing schemes to generate exploitable three-body interactions. 

In contrast, there recently has been a proposal to utilize existing anharmonicities in the Coulomb repulsion (or explicitly engineer such anharmonicities) in order to generate noise-resilient entangling gates for trapped ions \cite{Le2024}. In such cases, the subleading interaction terms are unwanted error sources, whose magnitudes need to be carefully considered.

From our analysis, we have seen that anharmonicities in the potential energy can induce two types of terms that need careful attention. The first one is a phonon-induced shift of the qubit resonance frequency---an effect that in the literature has been rarely discussed~\cite{Loewen2003}. Its strength increases with chain length and phonon occupation numbers. Although its mean effect can be compensated by appropriate detuning of the microwave radiation, phonon number fluctuations remain problematic for precise operations, thus further highlighting the need for efficient cooling techniques. The second effect that requires attention is a two-to-one conversion of phonons~\cite{Marquet2003}. It vanishes if all phonon modes are ground-state cooled, and it can be neglected in a rotating wave approximation if the trap is designed such that resonances between phonon frequencies are avoided. 
These potential effects need to be kept in mind in particular when scaling to large ion chains. 

Our analysis provides a guideline for the precision necessary in engineering trapped-ion quantum hardware. 
Future work along this direction may include more detailed investigations of the radial modes (see App.~\ref{app:transversal}), which can become coupled to the axial modes by anharmonic potentials, as well as expansions to even higher order (e.g., fourth-order correction to the Coulomb potential). 
Similar analyses may also be important for other approaches, e.g., those based on M{\o}lmer--S{\o}rensen-type gates using laser beams \cite{Moelmer1999,Soerensen1999} or oscillating magnetic fields \cite{Ospelkaus2008, Ospelkaus2011, Yu2022,Loeschnauer2024}, and they may also be important for other platforms such as superconducting qubits, which are governed by very similar microscopic Hamiltonians \cite{Blais2007}.
Finally, 
our study may inspire research into whether it is possible to purposefully enhance higher-order effects for the design of new quantum gates or the use in other quantum-information processing tasks.

\begin{acknowledgments}
We acknowledge useful discussions with Ivan Boldin, Nina Megier, Sebastian Rubbert and Theerapot Sriarunothai.
The work reported in this publication is based on a project that was funded by the German Federal Ministry for Education and Research under the funding reference number 13N16437. The authors are solely responsible for the content of this publication.
This work has benefited from Q@TN, the joint lab between University of Trento, FBK—Fondazione Bruno Kessler, INFN—National Institute for Nuclear Physics, and CNR—National Research Council. 
We acknowledge support by Provincia Autonoma di Trento. 

This Accepted Manuscript is available for reuse under a CC BY-NC-ND licence after the 12 month embargo period provided that all the terms and conditions of the licence are adhered to.
\end{acknowledgments}

\appendix

\section{Higher-order corrections from transversal phonon modes} \label{app:transversal}

In this appendix, we briefly discuss the corrections due to transversal phonon modes in an expansion of the potential energy up to third order. 
Although the transversal modes also cause corrections in a third-order expansion of the resonance frequencies of the ions, the corresponding terms are proportional to the curvature of the magnetic field in axial and radial directions. As argued in Sec.~\ref{subsec:higherorderB}, such contributions can safely be neglected for realistic magnetic fields.
We therefore focus here only on anharmonicities arising from the Coulomb potential. 

For a one-dimensional ion chain in a harmonic trap, analogously to the discussion in Sec. \ref{subsec:coulomb}, the external potential can be expanded around the equilibrium position $\bm{q}_0$ as \cite{Marquet2003}
\begin{align} \label{eq:transversal_coulomb}
    V(\mathbf{q})  &= V(\bm{q}_0) + V^{(2)}(\bm{q})+ V^{(3)}(\bm{q}) + \bigO(\vec{q}^4) \nonumber \\
    &= V(\bm{q}_0) + \frac{1}{2} \sum_{\alpha = 1}^3\sum_{i,j} A^{\alpha}_{ij} q_i^{(\alpha)} q_j^{(\alpha)}\nonumber \\
    &\hphantom{={}}+ \sum_{i,j,k} \Tilde{B}_{ijk} q_k^{(3)} \left(2 q_i^{(3)} q_j^{(3)} - 3 q_i^{(2)} q_j^{(2)} - 3 q_i^{(1)} q_j^{(1)} \right) \nonumber \\
    &\hphantom{={}}+ \bigO(\vec{q}^4) \,,
\end{align}
where $q_i^{(3)}$ ($q_i^{(2)}$, $q_i^{(1)}$)  is the displacement of ion $i$ from its equilibrium position in the axial (transversal) direction. Further, we have defined
\begin{align}
     A^\alpha_{ij} &= \left. \frac{\partial^2 V}{\partial q_i^{(\alpha)} \partial q_j^{(\alpha)}} \right|_{\bm{q}_0} \,, \\
     \Tilde{B}_{ijk} &= \frac{m \omega_z^2}{2l} \sum_{q \neq i} \frac{\text{sgn}(u_q - u_i)(\delta_{ij} - \delta_{qj})(\delta_{ik} - \delta_{qk})}{(u_q - u_i)^4} \,, 
\end{align}
with the ion mass $m$, the axial trap frequency $\omega_z$, the characteristic length scale $l$ (defined as $l^3 = e^2/ 4\pi \epsilon_0 m \omega_z^2$), and the dimensionless equilibrium positions $\bm{u} = \bm{q}_0/l$.

The second-order terms in Eq.~\eqref{eq:transversal_coulomb} are decoupled. After quantizing and performing a polaron transformation on these terms with respect to all axial and transversal modes, similarly as in Sec.~\ref{sec:magic}, we arrive again at Eq.~\eqref{eq:magic_coupling}, where the spin--spin coupling strength is now given by 
\begin{align}
    J_{ij}^{(2)} &= \sum_{\alpha = 1}^3\sum_{n=1}^N \nu_n^{(\alpha)} \epsilon^{(\alpha)}_{in} \epsilon^{(\alpha)}_{jn} \,.
    \label{eq:J2_transversal}
\end{align}
Here, we have defined the $n$-th phonon mode frequency $\nu_n^{(\alpha)}$ in direction $\alpha = 1, 2, 3$ and the effective Lamb--Dicke parameter $\epsilon^{(\alpha)}_{in}$ describing the coupling of ion $i$ to said mode, cf.\ Eq.~\eqref{eq:lambdicke}. The coupling strength scales as $J_{ij}^{(2)} \sim \sum_\alpha (\partial_\alpha B / \nu^{(\alpha)})^2$, where $\partial_\alpha B$ is the magnetic field gradient in direction $\alpha$. If the magnetic gradient vanishes in both transversal directions, the spin--spin coupling strength reduces to Eq.~\eqref{eq:J2} stated in the main text.

For the third order terms in Eq.~\eqref{eq:transversal_coulomb} (third row), we find one term with only axial motion ($\sim q_k^{ (3)}q_i^{(3)} q_j^{(3)}$), which we have already discussed in Sec.~\ref{subsec:coulomb}, and two terms which couple axial and transversal motion ($\sim q_k^{ (3)}q_i^{(2)} q_j^{(2)}$ and $\sim q_k^{ (3)}q_i^{(1)} q_j^{(1)}$). For the latter two terms, we can perform the same analysis as for the first term (see Sec.~\ref{subsec:coulomb}).

If the magnetic gradient vanishes in the transversal directions, we are then left with two terms giving corrections to the local fields and to the three-body mode couplings, respectively.
The first one is given by 
\begin{subequations}
\begin{gather}
\begin{split}
    \sum_{\alpha = 1}^2\sum_{ijk} C_{ijk}^\alpha \big(&a_{\alpha , j}^{\dagger} a_{\alpha ,k}^\dagger 
    + a_{\alpha ,j} a_{\alpha ,k} \\
    &+ a_{\alpha ,j}^\dagger a_{\alpha ,k} + a_{\alpha ,j} a_{\alpha ,k}^\dagger\big)
    \sum_n \epsilon_{ni}^{(3)}\sigma_Z^{(n)} \,, \\
\end{split}
    \\
    C_{ijk}^\alpha = -3 \Delta z_i \Delta \alpha_j \Delta \alpha_k \sum_{mnp} \Tilde{B}_{mnp} S_{pi}^{(3)} S_{mj}^{(\alpha)} S_{nk}^{(\alpha)} \,, 
\end{gather}
\end{subequations}
with $S^{(\alpha)}$ being the mode participation matrix for direction $\alpha$, and $\Delta \alpha_j = \sqrt{\hbar / 2m\nu_j^{(\alpha)}}$. Similar to the discussion in Sec.~\ref{subsec:coulomb}, terms that do not preserve phonon energy can be neglected within a rotating-wave approximation. The remaining corrections to the local fields read 
\begin{align}
    \sum_n \sigma_Z^{(n)}\sum_{\alpha=1}^2 \sum_{ik}  C_{iik}^\alpha  \epsilon_{ni}^{(3)} \left(2 a_{\alpha,i}^{\dagger} a_{\alpha, i} +1 \right) \,.
\end{align}
This correction scales as $\sim \sum_{\alpha=1}^2\partial_3 B / \nu^{(\alpha)} \nu^{(3)}$. Comparing this to the correction solely due to the axial motion (see Sec.~\ref{subsec:coulomb}), which scales as$\sim \partial_3 B /  \nu^{(3)2}$, we can conclude that this contribution is smaller, as the trapping frequencies in the transversal directions are typically much larger than the axial trap frequency \cite{Piltz2016, Jurcevic2014}. While the mean contribution of this effect, like for the axial modes discussed in \cref{subsec:coulomb}, can be easily compensated, the fluctuations in the transversal modes, despite having a smaller effect, can still affect gate fidelity, if the system is not sufficiently cooled down.

The only other non-vanishing terms in Eq.~\eqref{eq:transversal_coulomb} (for vanishing magnetic field gradients in the transversal directions) are the three-body axial-transversal mode couplings
\begin{align}
    \sum_{\alpha=1}^2 \sum_{ijk}  C_{ijk}^\alpha \big(a_{3 ,k}^\dagger + a_{3 ,k}\big) \big(a_{\alpha ,i}^\dagger + a_{\alpha ,i}\big) \big(a_{\alpha ,j}^\dagger + a_{\alpha ,j}\big) \,.
\end{align}
Depending on the trap frequencies and system size, resonances could arise in some of these interaction terms \cite{Marquet2003}. Although they would then remain relevant even after a rotating wave approximation (see discussion in Sec.~\ref{subsec:coulomb}), they can still be avoided by  cooling the system sufficiently close to the ground state.

\bibliography{trappedions2}

\end{document}